# Toward the Integration of Traditional and Agile Approaches

Hung-Fu Chang
Computer Science
University of Southern California
Los Angeles, United States
hungfuch@usc.edu

Stephen C-Y. Lu
Viterbi School of Engineering
University of Southern California
Los Angeles, United States
sclu@usc.edu

*Abstract*—The agile approach uses continuous delivery, instead of distinct procedure, to work closer with customers and to respond faster requirement changes. All of these are against the traditional plan driven approach. Due to agile method's characteristics and its success in the real world practices, a number of discussions regarding the differences between agile and traditional approaches emerged recently and many studies intended to integrate both methods to synthesize the benefits from these two sides. However, this type of research often concludes from observations of a development activity or surveys after a project. To provide a more objective supportive evidence of comparing these two approaches, our research analyzes the source codes, logs, and notes. We argue that the agile and traditional approaches share common characteristics, which can be considered as the glue for integrating both methods. In our study, we collect all the submissions from the version control repository, and meeting notes and discussions. By applying our suggested analysis method, we illustrate the shared properties between agile and traditional approaches; thus, different development phases, like implementation and test, can still be identified in agile development history. This result not only provides a positive result for our hypothesis but also offers a suggestion for a better integration.

## I. INTRODUCTION

Developing a modern software system becomes very challenging due to the increasing customer demands on more functions and higher quality. Especially, when software engineers face this challenge under very dynamic market, how to change their software or service in order to satisfy the need of faster delivery, better quality and lower cost rises many discussions [1, 2].

Recently, many suggestions for improving software development methods have come from real world practitioners. The main trend is the agile method. Unlike traditional development method, which requires a disciplined and distinct procedure, the agile development places the highest priority on satisfying the customer needs through continuous delivery [3, 4]. It emphasizes on rapidly iterations with the focus on working software so it can embrace closely customer collaboration by using faster responses to changing needs.

In addition, the theory behind the traditional methods is that all the requirements can be defined at the beginning of the system building process and a sequence of well-articulated tasks like systems planning, analysis, architecture, design, development, and testing can be explicitly defined [5]. Therefore, the development process is systematic, and the boundary of each task can be clearly identified. On the other hand, the agile method is more chaotic. It contains the evolutionary delivery through short iterative cycles that blending planning, action, and testing activities within intense human collaboration.

Software industry found that agile process fits small and stand-alone projects better. Developers and managers have difficulties to scale up and to integrate agile practices into the organization that already has well-defined traditional process. Therefore, industry seeks a solution of integrating agile and traditional methods so their benefits can be synthesized [18]. Past studies have discussed the agile method in the area of focusing on the integration of both traditional and agile developments or comparison of these two different methods [6]. Those suggested integration methods and the comparison studies are mostly inferred from the description of development activities or the review of the process. But, there are not any comparison research or any integration method, which has previously been published in the aspect of source code and design artifact's data analysis. Therefore, one shortcoming of these studies is lacking of supportive evidences from scientific data analysis. To remedy this, we would like to investigate how the agile method is executed in practices and what their results or effects look like. We argued that agile and traditional developments should still share many similar characteristics although the whole agile development could be chaotic due to putting various tasks together in a single iteration. Once we identify different phases, such as requirement defining, implementation, and testing in the whole agile development history, how to integrate traditional and agile methods or how to compare them can be further developed.

In this paper, we investigate the history of a software project, which is developed by the agile approach. By cross-referencing the source code and analysis of development log and meeting notes, we identify several characteristics of the agile development. We find that agile project development is not so chaotic. It still demonstrates systematic aspects, like the traditional software development.

The rest of this paper will be organized as follows. Section 2 will explore the related research. Section 3 will discuss the detailed differences between agile and traditional software developments. Section 4 will explain the analysis method. Section 5 will show the analysis results and then discuss them. Finally, we will conclude our research and explain our future research.

## II. RELATED WORK

Many past studies reveal the differences or contradictions between traditional and agile developments, tried to integrate both methods by applying the agile method to traditional approach. Parsons and Lai [7] discussed the hybrid approaches in the software quality perspective and argued the differences based on the statistics. Manhart and Schneider [8] showed the integration of agile and traditional methods an industrial case study. They claimed that both approaches shared the common developing goal but had different kinds of emphases. Armitage [9] described another hybrid approach that overlays the agile process with higher level design approaches in order to assist refactoring. Turner and Jain [10] researched the culture clash between the agile and Capability Maturity Model Integration (CMMI) processes. Lycett et al [11] suggested a situated process framework, in which, patterns are developed through a situated examination of contextual characteristics (e.g., project, product, or team)and expressed as Rational Unified Process (RUP) development cases. Alegria and Bastarrica [12] discussed the way to reach CMMI level 2's certification by implementing agile methods like Scrum and Extreme Programming (XP).

Several previous reviews were also published to introduce characteristics of the agile method by comparing both agile and traditional approaches. Cohen et al.'s [13] explored the history of agile development, and particularly discusses relations between agile development and the Capability Maturity Model (CMM). Wang et al. introduced the contradictions in the agile development and used a paradoxical perspective to deal with them. Nerur et al. explored the differences and pointed out the challenges of changing to the agile method.

Most past research proposed their integrated approach by inserting the agile method into traditional development because their assumption is that the developer can treat traditional approach as an outline and then add the agile activities inside each major phase. However, the validation of this type of study lacks of the perspective of the data analysis about the delivered artifacts. With the implementation data analysis supports, the differences between agile and traditional development can become clearer and both methods' benefits can be synthesized seamlessly.

## III. AGILE AND TRADITIONAL SOFTWARE DEVELOPMENT

### A. *Agile vs Traditional software Development*

One major reason to cause the failure of a software project is that the built software system cannot be delivered on time. Even if the software can be delivered on time, it may not satisfy all the customer's expectations. As a result, agile software development is created to solve these problems. However, agile methods also face some critics, for example, insufficient architecture planning, over-emphasis on early results, and low levels of test coverage. These shortages can also be explicitly observed and understood while two development process models are compared.

In the traditional software development, each step in the process is clear. One must start only after the previous step is completed. On the other hand, software engineers who use the agile development do not wait for prior procedure to complete (see Fig. 1). Each iteration, engineers review their results, and then modify and test the product in the next iteration.

### B. *Observation of Source Code Changes*

The implementation in traditional software process usually starts after a thought-through design. The amount of source codes usually increases largely during the early phase of development because most function has been implemented. After the main structure of the system becomes steady, the lines of source codes gradually increase or decrease. Thus, the source codes in the traditional method do not have dramatic dynamic changes (i.e., rapidly increase or decrease in a large amount) at anytime in the whole development.

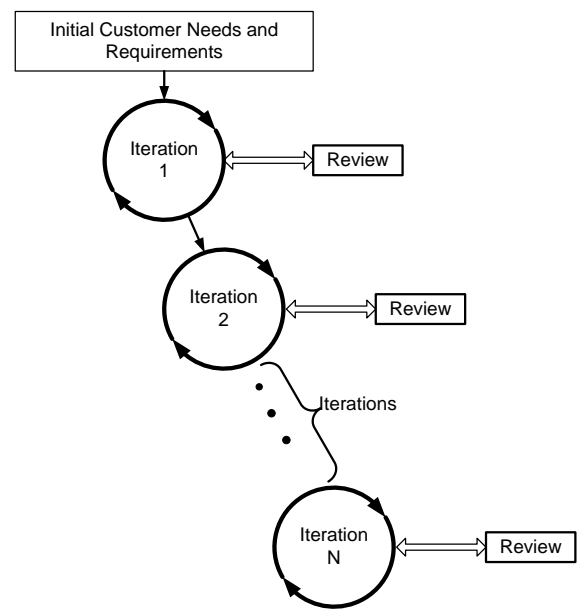

Fig. 1. Agile Software Development Process

However, the agile method shows very differently. The key characteristic of the agile method is rapid iteration. After every time's iteration, the release tries to meet customer's requirements. If not, changes should immediately happen in the next iteration. The amount of source code change depends on customer's review result. Due to no solid development planning, it is very possible to have changes about major structure adjustment. In addition, requirements during the whole development could change very often. As a result, the source codes change largely. However, we argue that the agile development can also demonstrate certain similar characteristics as traditional development. In our analysis, we would like to investigate this observation.

## IV. ANALYSIS

### A. Scope of Application

Our method targets the analysis of agile development project. The agile development team's software release and team meeting is weekly. Although primary software releases and weekly meetings are stored, between two weekly releases, there may still many development versions committed in the repository as well as many discussions, and documents are saved. Therefore, our analysis method will be applied to these data.

### B. Data Analysis Method

There are three major stages in our analysis. We firstly collect data from the agile development project. In second stage, we eliminate insignificant versions from our collected dataset in order to reduce the efforts of the analysis. Lastly, we identify those phases, such as requirement, software architecture, implementation, or testing, as we define in the traditional development. Finding these phases is the key step in our justification of our source analysis hypothesis.

*1) Data collection and engineering*

Three types of data are collected from the project: meeting notes, source codes, and version logs. Two programs are written for collecting those data. The first program extracts all the source codes and version logs from Subversion (SVN) version control repository. Since the weekly meeting notes are written in the MS Power Point or Word formats and discussions are posted the internal wiki website. These textual data are first extracted by the other program and then are reorganized in to a time series structure. Using this time series structure can help us to specify various phases according to the project's development timeline.

*2) Identify key versions and development task*

Because many versions are only saved for records, their modifications are small and cannot reflect structure altering, important designer's decisions, or requirement changes. To avoid analyzing these trivial versions, one of our jobs is to identify the key versions in the development history. We use two ways to identify key versions - source code and text analyses.

*a) Source code analysis*

One characteristic of key versions in the source code analysis is that the amount of code change is substantial. Therefore, by comparing the number of source line of code (SLOC) in two sequential versions, those key versions can be identified. More importantly, in a source analysis diagram (e.g., SLOC VS version), the key version points can match the shape of the curve and capture the turning points.

To determine the key versions, we develop three methods to extract those versions that match significant changes. The first method calculates the slop change (*SC*) against three consecutive versions (see Fig. 2).

$$SC = (V_{n+1} - V_n) / (V_n - V_{n-1}) \quad (1)$$

$V_n$ is the measured value (e.g., SLOC in the SLOC VS version diagram) at version n, $V_{n+1}$ represents is the measured value at version $n+1$, and $V_{n-1}$ represents is the measured value at version *n-1*.

This *SC* represents the angle between two tangents from two sequential versions. Once the *SC* exceeds the threshold, the key versions can be extracted.

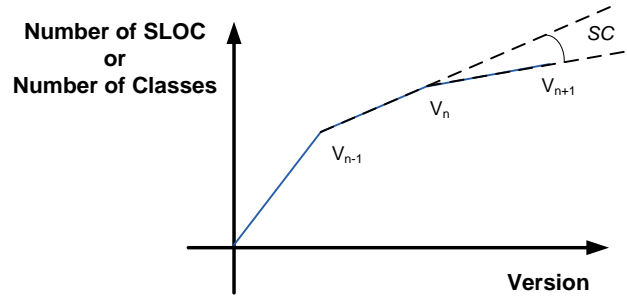

Fig. 2. The Calculation of the Slope Change

The second method extracts key versions based on the calculation of the relative difference *RC* between two sequential versions (see Fig. 3). We can also setup a threshold to determine if a version is the key version. The equation below is to calculate the relative changes *RC*.

$$RC = (V_n - V_{n-1}) / V_{n-1} \quad (2)$$

$V_n$ is the measured value at version *n* (e.g., SLOC in the SLOC VS version diagram), and $V_{n-1}$ represents is the measured value at version *n-1*.

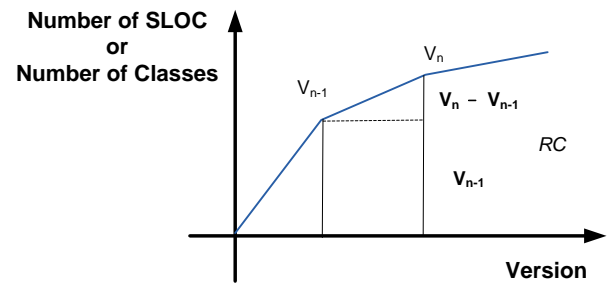

Fig. 3. The Calculation of the Relative Change

The third method is very similar to the second one. Instead of calculating the relative change, we compute the direct difference based on a normalized curve. The normalized values are calculated based on the measured value divided by the maximum value; for example, each number of SLOC divided by maximum number of SLOC. After we get the normalized values, we can direct calculate the difference using the formula below.

$$DC = (V_n - V_{n-1}) \quad (3)$$

*DC* is direct change, $V_n$ is the normalized value at version *n*, and $V_{n-1}$ represents the normalized value at version *n-1*.

To avoid missing any significant modifications or possible key version, the union of all the above three method's results is the entire key version set.

*b) Text analysis*

We applied text analysis in the developer's meeting notes and version logs. Since text analysis could be very time consuming, rather than analyzing every version's log, only key version's log is used. This method firstly detects the keywords according to its frequency, developer's descriptive guideline, or common terminologies. For example, in the log "fixed the bug no 23", "bug" and "fix" can be two keywords which represent correcting the program to satisfy the functional requirement. Then, we manually identify the description about requirements or development planning from meeting notes. After we analyze the meeting notes and version logs, we can decide which type of development task, such as debugging, building new function, or testing, is the major activity between two versions.

*3) Identify different phases*

To identify different phases, we need to do cross-referencing between the result of source code and text analysis. The source code analysis tells us that which versions are representative in the whole version history. The text analysis shows the type of development activity between two versions. With combining these two kinds of outcomes, we can further understand the major development activity within a period of time. In addition, the meeting note analysis result can also be used to verify if the phase that we identify is valid or not.

V. RESULTS AND DISCUSSIONS

A. *Case Study Project Background*

The agile project that we investigate is called Visualization of Attack Surfaces for Targeting (VAST). This is a tool that is developed based on the Eclipse plug-in framework. The VAST tool provides multi-column code viewer with bread crumb trail so that it helps code auditors to retrace their thought processes and shows source code overview in context of the software vulnerability. The tool is developed by a five developer's product team in the Information Sciences Institute. The entire developing time is 18 months, and 841 versions are committed.

The team follows many methods that the agile practices proposed during entire development. First, their customer worked closely with the team, like one of the team members. The customer immediately clarifies their needs and identifies the priority. The team does not hold any meetings to layout the whole system structure; instead, customers reveal the expected user interfaces. Second, the tool is released weekly, an acceptance test is applied, and the customer discusses the expectation in the next release. Third, the team has daily meet like scrum to know each member's obstacles and status. Lastly, the team keeps refactoring the code. In addition, team members also use an internal wiki site to maintain all the documents, discussions, learning, and meeting notes. Since there is a software release every week, in the initial stage of the project, the VAST team does not spend much time to work on the software architecture; instead, they quickly divide the task and start to build the software. The team expects the software will finally change while they have better understanding on design and customer's needs after several iterations.

B. *Key Version Extraction*

We apply equation (1), (2), and (3) to all the version history in order to identify those key versions. In equation (1), (2), and (3), we use 0.2, 0.2, and 0.15 as thresholds, respectively. The entire key version is the union set of the results of all these three equations. In Fig. 3 and 4, we pick those key version points on the both original SLOC and Number of Classes curves, respectively Then, we connect all the points to form a curve that matches the original graph. The matching curves in both Fig. 4 and 5 obviously still reserve the characteristics of the original curves. This shows that the matching curves should be able to have enough significance to represent the original curves. Therefore, we have confidence to use these key versions to do our next step analysis.

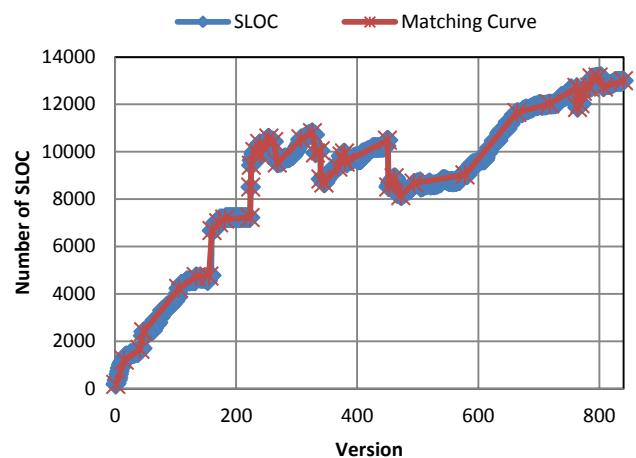

Fig. 4. Version VS SLOC and Its Matching Curve

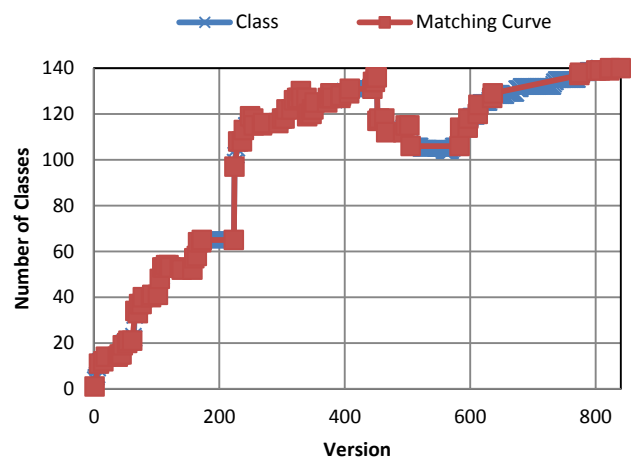

Fig. 5. Version VS Number of Classes and Its Matching Curve

Fig. 6 shows the normalized matching curves of Number of Classes and Number of SLOC. All the versions in Fig. 6 are potentially the points that separate two different phases. Therefore, by investigating the meeting notes, documents, and discussions, we can validate separation points and detect different type of development stage. In our text analysis, except the logs of task assignment (e.g., developer X should work on task 1), three kinds of descriptions can be identified. They are functional, modification and testing or debugging descriptions. The log also tells the re-factored versions that are those sharp change points in matching curve. As well, in the meeting notes, we can find when customers stop to request modification of the system due to the stabilized needs. By knowing this time points and various types of descriptions in the logs, through a cross-reference between discussion, logs, and meeting notes, we can be divided the whole development history into four phases: (1) customer needs to requirements (from version 1 to 173) (2) developer's learning and research (from version 174 to 264) (3) implementation and testing (from version 264 to 625) (4) debugging (from version 626 to the end of development - version 841).

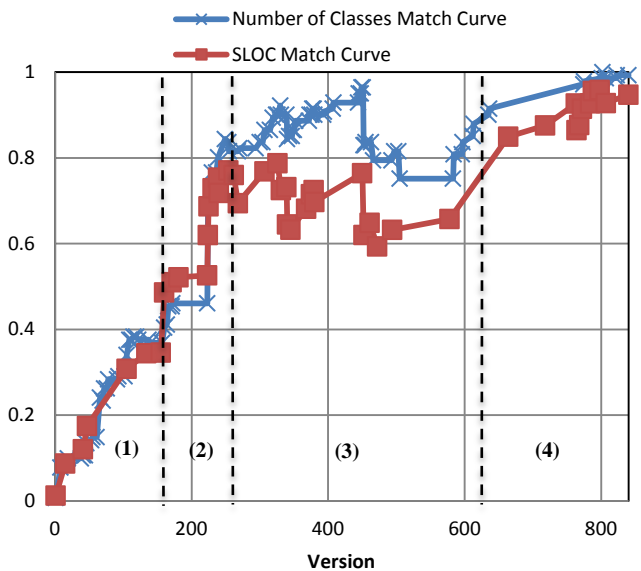

Fig. 6. Normalized Matching Curves of Version VS Number of Classes and Version VS SLOC

Moreover, Fig. 6 also implies that the functional change should be less or adding function is completed when the development reaches a point where the whole system and requirements are more stable (i.e., the end of the second phase). After this point, the development activities turn to be focusing on testing and debugging.

## VI. CONCLUSION AND FUTURE RESEARCH

The agile approach recently becomes a main trend in both industry and academia. Due to this, many studies try to understand the differences between this new and old development approaches to gain a balance between them, and then the benefits of these two methods can be synthesized. While there is no concrete data analysis of the implementation to support the integration of both methods from past research, our research particularly uses source code and design artifact analysis to complement the type of study.

In our source analysis method, we capture the characteristics of the SLOC VS version curve and then using this normalized skeleton curve to specify each development phase as traditional plan driven approach. From our case study result, we find that agile and traditional approaches share common features. The agile development has distinguished each phase like traditional process. This provides a data analysis evidence of the integration. We discover that the agile activities can be treated as the sub-activities in primary development phase. Our research may lack of the suggestion from management's point of view but we do provide another perspective to the agile approach.

Since our paper only contains one project data analysis, in the future, we should collect multiple projects' data in order to strengthen our conclusion. In addition, we may also apply our source code analysis to enhance the software process improvement so that the integration can be more precise and seamless. Particularly, we would like to further research about offering good advices for managing a software project that could adopt the agile method.